\documentclass[pra,showpacs,floatfix]{revtex4}
\usepackage{amssymb}
\usepackage{graphicx}

\begin{document}

\title{Stable periodic density waves in dipolar Bose-Einstein condensates
trapped in optical lattices}
\author{ Aleksandra Maluckov$^{1}$, Goran Gligori\'{c}$^{2,3}$, Ljup\v co Had%
\v{z}ievski$^{3}$, Boris A. Malomed$^{4}$, and Tilman Pfau$^{5}$}
\affiliation{$^1$Faculty of Sciences and Mathematics, University of Ni\v s, P. O. B. 224,
18000 Ni\v s, Serbia \\
$^2$ Max-Planck-Institut f\"ur Physik komplexer Systeme, N\"othnitzer Stra%
\ss e 38, D-01187 Dresden, Germany \\
$^3$ Vin\v ca Institute of Nuclear Sciences, University of Belgrade, P. O.
B. 522,11001 Belgrade, Serbia\\
$^{4}$ Department of Physical Electronics, School of Electrical Engineering,
Faculty of Engineering, Tel Aviv University, Tel Aviv 69978, Israel\\
$^{5}$Physikalisches Institut, Universit\"{a}t Stuttgart, Pfaffenwaldring
57, 70569 Stuttgart, Germany}

\begin{abstract}
Density-wave patterns in (quasi-) discrete media with local interactions are
known to be unstable. We demonstrate that \emph{stable} double- and triple-
period patterns (DPPs and TPPs), with respect to the period of the
underlying lattice, exist in media with nonlocal nonlinearity. This is shown
in detail for dipolar Bose-Einstein condensates (BECs), loaded into a deep
one-dimensional (1D) optical lattice (OL), by means of analytical and
numerical methods in the tight-binding limit. The patterns featuring
multiple periodicities are generated by the modulational instability of the
continuous-wave (CW) state, whose period is identical to that of the OL. The
DPP and TPP emerge via phase transitions of the second and first kind,
respectively. The emerging patterns may be stable provided that the
dipole-dipole (DD) interactions are repulsive and sufficiently strong, in
comparison with the local repulsive nonlinearity. Within the set of the
considered states, the TPPs realize a minimum of the free energy.
Accordingly, a vast stability region for the TPPs is found in the parameter
space, while the DPP\ stability region is relatively narrow. The same
mechanism may create stable density-wave patterns in other physical media
featuring nonlocal interactions, such as arrayed optical waveguides with
thermal nonlinearity.
\end{abstract}

\maketitle

\textit{Introduction. }The formation of density-wave patterns in
periodically structured media is a fundamental physical effect, celebrated
manifestations of which are the Peierls instability of the electron gas in
one-dimensional (1D) metals, leading to the emergence of charge-density
waves \cite{Peierls,spin}, and spin-density waves \cite{spin}. Recently,
many complex settings known in solid-state physics have been reproduced (%
\textit{simulated}) in an essentially simpler form in quantum gases---often,
these are atomic Bose-Einstein condensates (BECs)---loaded into optical
lattices (OLs) \cite{simulator}. Thus far, this possibility was not
demonstrated for density-wave patterns, the reason being that such patterns
formed in BEC with contact interactions between atoms are unstable \cite{dps}%
. The objective of this work is to demonstrate that \emph{stable} density
waves are possible in BEC with long-range dipole-dipole (DD) inter-atomic
interactions. The same mechanism should make it possible to create stable
density (intensity) waves in other periodically structured physical media
featuring a nonlocal nonlinear response, which is inherently present in the
transport mechanisms of heat \cite{thermal} and charge carriers \cite{semi},
electrostatic interactions in liquid crystals \cite{liq}, photon
self-attraction \cite{photon}, and many-body interactions in plasmas \cite%
{plasma} and BEC \cite{matter}. Further, an effective gravitation-like
attraction between atoms can be optically induced in BEC by means of
resonant illumination \cite{grav}. It was demonstrated theoretically \cite%
{Krolik} and experimentally \cite{long} that the implementation of patterns
supported by the nonlocal nonlinearity is especially relevant in thermal
optical media, where the periodic structure can be readily built in the form
of arrayed waveguiding systems.

BEC of dipolar atoms or molecules loaded into OLs have been in the focus of
research work for the past decade, starting from the theoretical analysis of
basic properties of such condensates \cite{theory}, followed by the analysis
of quantum phases \cite{quantum-phases-in-OL,quantum-phases}, various
textures \cite{textures,triple}, and suppression of the quantum collapse
(fall onto an attractive center) \cite{fall} in them. Structured ground
states and supersolidity of dipolar gases \cite{theory,supersol2,Stuttgart},
as well as the rotonic dispersion relation \cite{rotonic}, were discussed
theoretically but still await experimental confirmation. Experimentally,
first effects of dipolar interactions in quantum gases were observed in the
BEC of $^{52}$Cr atoms \cite{experiment}. A condensate of $^{164}$Dy atoms
was recently created too \cite{Dy}, which provides an even stronger DD
interaction than $^{52}$Cr. In addition to that, the creation of dipolar BEC
is expected in erbium \cite{mcclelland} and in gases of molecules carrying
electric dipole moments \cite{hetero}. A powerful tool which allows one to
control dynamical effects in the dipolar BEC, such as the onset of collapse,
is the use of the Feshbach resonance for tuning the strength of contact
interactions between atoms, which compete with their long-range DD
interactions \cite{nature}. The results obtained in this field have been
summarized in recent review \cite{ddibec}.

Another ubiquitous tool used in experiments with dipolar BEC is provided by
OLs. Recently, dipolar effects in the chromium condensate trapped in OLs
were reported in Refs. \cite{France,new,fatori}. In particular, strong DD
repulsion can stabilize the condensate with attractive contact interactions,
trapped in the OL \cite{new}. It is well known that the OL facilitates the
creation of gap solitons \cite{gap}, or bound pairs of repulsively
interacting atoms, at the microscopic level \cite{nature1}. Gap solitons in
the 1D model of dipolar condensates trapped in the OL potentials were
studied in Ref. \cite{Cuevas}.

While the state of the trapped condensate may obviously feature
the same periodicity as the underlying lattice (we call it the
continuous-wave state, CW), we here report \emph{stable} density
waves in the form of double- and triple-period patterns (DPPs and
TPPs) in the dipolar BEC loaded into a deep OL potential. The TPP
realizes a relatively deep energy minimum in the space of the
considered states, and its stability area is much broader than
that of the DPP, therefore the TPP is a candidate to the role of
the ground state. To the best of our knowledge, this is the first
demonstration of stable density waves of the TPP type in
OL-trapped BECs, as well as in dynamical lattices with long-range
interactions of any physical nature. Featuring long-range
off-diagonal and diagonal order different from that of the
underlying lattice potential, both the DPPs and TPPs may be viewed
as a metastable supersolid state
\cite{supersolid,supersol2,Stuttgart} of the dipolar condensate.

In Ref. \cite{dps} it was demonstrated that the period-doubling instability
of the CW in the BEC with local interactions, trapped in the deep OL, gives
rise to DPPs, but, as mentioned above, the emerging pattern is \emph{never}
stable. As for the TPPs, an incentive for their consideration is provided by
the recent analysis of the trapped dipolar BEC, which concluded that a
three-well system is the minimal setting which is necessary to let the
nonlocal character of the DD interactions manifest itself \cite{triple}. In
this work we consider all possible combinations of the repulsive/attractive
contact (RC/AC) and repulsive/attractive DD (RDD/ADD) interactions, and
demonstrate that \textit{unstaggered} (regular) DPPs and TPPs have their
\emph{stability regions} in the RC+RDD case, provided that the long-range DD
repulsion is sufficiently strong in comparison with the local self-repulsion
(these findings imply that, in the opposite AC+ADD case, the \textit{%
staggered} \cite{Panos} counterparts of these patterns are stable in the
respective regions too). The analysis reveals that the multiple-period
patterns emerge when the corresponding modulational instability of the CW
sets in, i.e., the respective perturbations seed the creation of the
patterns. The TPP's stability area is found to be much larger than that of
the DPP, and the calculation of the free-energy density demonstrates that
the TPP realizes a well-pronounced energy minimum, in comparison with the CW
and DPP states. When the TPPs are unstable, the DD interactions of either
sign significantly inhibits the instability, suggesting that even unstable
TPPs may be observed in the experiment.

\textit{The model}. For condensates loaded into a deep OL, the corresponding
Gross-Pitaevskii equation can be reduced to the discrete nonlinear Schr\"{o}%
dinger equation (DNLSE) in the framework of the tight-binding approximation
\cite{DNLS,dps} (a similar reduction to the quasi-discrete propagation is
well known in optics \cite{review}). In turn, the analysis of various
stationary states supported by the DNLSE may start from the anti-continuum
limit, which corresponds to the chain of uncoupled sites \cite{anti}. The
same tight-binding approximation, if applied to the dipolar BEC loaded into
a deep OL, leads to the 1D \cite{quantum-phases-in-OL,nashi} and 2D \cite%
{takode-nashi} DNLSEs which include the long-range DD interaction between
lattice sites. We here consider the trapped condensate modeled by the scaled
1D equation \cite{nashi}:
\begin{equation}
i\frac{df_{n}}{dt}=-C(f_{n+1}+f_{n-1}-2f_{n})+\sigma \left\vert
f_{n}\right\vert ^{2}f_{n}-\Gamma \sum_{m\neq n}\frac{\left\vert
f_{m}\right\vert ^{2}}{|m-n|^{3}}f_{n},  \label{eq1}
\end{equation}%
where $\sigma =-1$ and $+1$ corresponds to the attractive and repulsive
contact interaction, respectively, $C>0$ is the inter-site coupling constant
which accounts for the tunneling of atoms between adjacent sites of the
lattice, and $\Gamma >0$ or $\Gamma <0$ is the relative strength of the DD
attraction or repulsion, in comparison to the local nonlinearity. In fact, $%
\sigma $ accounts for the \textit{effective} on-site nonlinearity, which
includes both the DD interaction between atoms trapped in a given potential
well and the contact interaction proper. Time is measured in units of $%
\omega _{\bot }^{-1}$, where $\omega _{\bot }$ is the frequency which
defines the radius of the transverse confinement, $a_{\bot }=\sqrt{\hbar
/(m_{\mathrm{atom}}\omega _{\bot })}$. $C$ may also be fixed by rescaling,
therefore basic findings are displayed below for $C=0.8$. In this work, we
present results for unstaggered patterns. They can be made equivalent to
their staggered counterparts by transformation $f_{n}\rightarrow
(-1)^{n}e^{4iCt}f_{n}^{\ast }$, $\left\{ \sigma ,\Gamma \right\} \rightarrow
-\left\{ \sigma ,\Gamma \right\} $.

Equation (\ref{eq1}) does not include the external trapping potential. Under
realistic experimental conditions, the actual number of OL periods in the
trap is $\gtrsim 100$. It is straightforward to check that distortion
introduced by this potential is negligible for the patterns reported below.
Undistorted configurations can also be realized in the dipolar condensate
loaded into a toroidal trap \cite{torus}.

For other nonlocal systems, such as optical media with the thermal
nonlinearity \cite{Krolik}, the DNLSE differs by the form of the kernel
accounting for the long-range interaction. Essentially the same analysis as
reported below can be developed for such models too.

\begin{figure}[h]
\center\includegraphics [width=7cm]{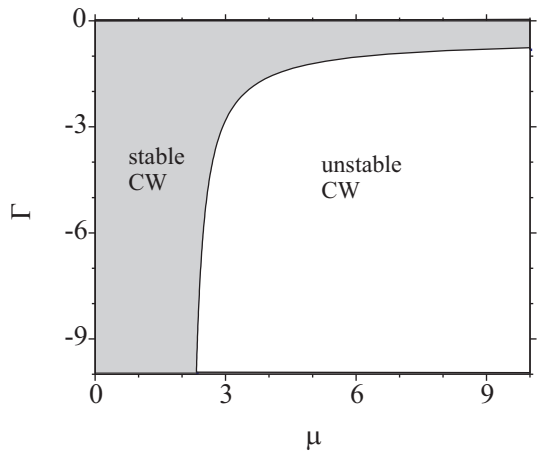}
\caption{The stability diagram for unstaggered CW states with the repulsive
sign of both the contact and DD interactions ($\protect\sigma =+1,\Gamma <0$%
). In all figures, the lattice coupling constant is $C=0.8$.}
\label{slikami}
\end{figure}

Stationary solutions to Eq. (\ref{eq1}) with chemical potential $\mu $ are
sought for as $f_{n}=U_{n}\exp {(-i\mu t)}$, which leads to the equation for
real discrete wave function $U_{n}$:
\begin{equation}
\mu U_{n}=-C(U_{n+1}+U_{n-1}-2U_{n})+\left[ \sigma U_{n}^{2}-\Gamma
\sum_{m\neq n}\frac{U_{m}^{2}}{|m-n|^{3}}\right] U_{n}.  \label{eq3}
\end{equation}%
First, a straightforward analysis of the modulational instability of the
unstaggered CW (constant) solution, $U_{n}^{2}\equiv \mu /\left[ \sigma
-2\zeta (3)\Gamma \right] $, where $\zeta (3)\approx \allowbreak 1.202$ is
the zeta-function, has been performed, using the linearization of Eq. (\ref%
{eq1}) for small perturbations. In the most relevant RC+RDD case, the CW is
stable in a relatively narrow parameter region, see Fig. \ref{slikami}.

\textit{Stable double-period and triple-period patterns}. The formation of
multiple-period patterns is a consequence of the modulational instability of
the CW states, which represent uniform chains of BEC droplets trapped in the
OL. Accordingly, periodically modulated patterns may be called \textit{%
condensate-density waves}. They were found numerically by solving Eqs. (\ref%
{eq3}) with the help of the modified Powell method \cite{nashi}, and also in
exact and approximate analytical forms (see below). Their stability was
explored by computing eigenvalues for small perturbations, using the
linearization of Eq. (\ref{eq1}), which was then checked by direct
simulations of Eq. (\ref{eq1}) for perturbed solutions. Close to the
anticontinuum limit, patterns with all periodicities are long-lived ones, as
the instability weakens with the decay of the inter-site coupling. This is a
generic property of the DNLSE with the contact nonlinearity \cite{Panos},
which persists in the presence of the DD interaction.

\begin{figure}[th]
\center\includegraphics [width=7cm]{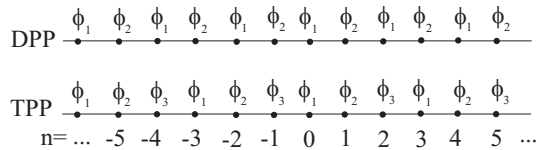}
\caption{A scheme of the condensate density waves with the double and triple
periodicity. $\protect\phi _{1,2,3}\,$ are amplitudes of the wave function
at the lattice sites.}
\label{slika1}
\end{figure}

\begin{figure}[th]
\center\includegraphics [width=13cm]{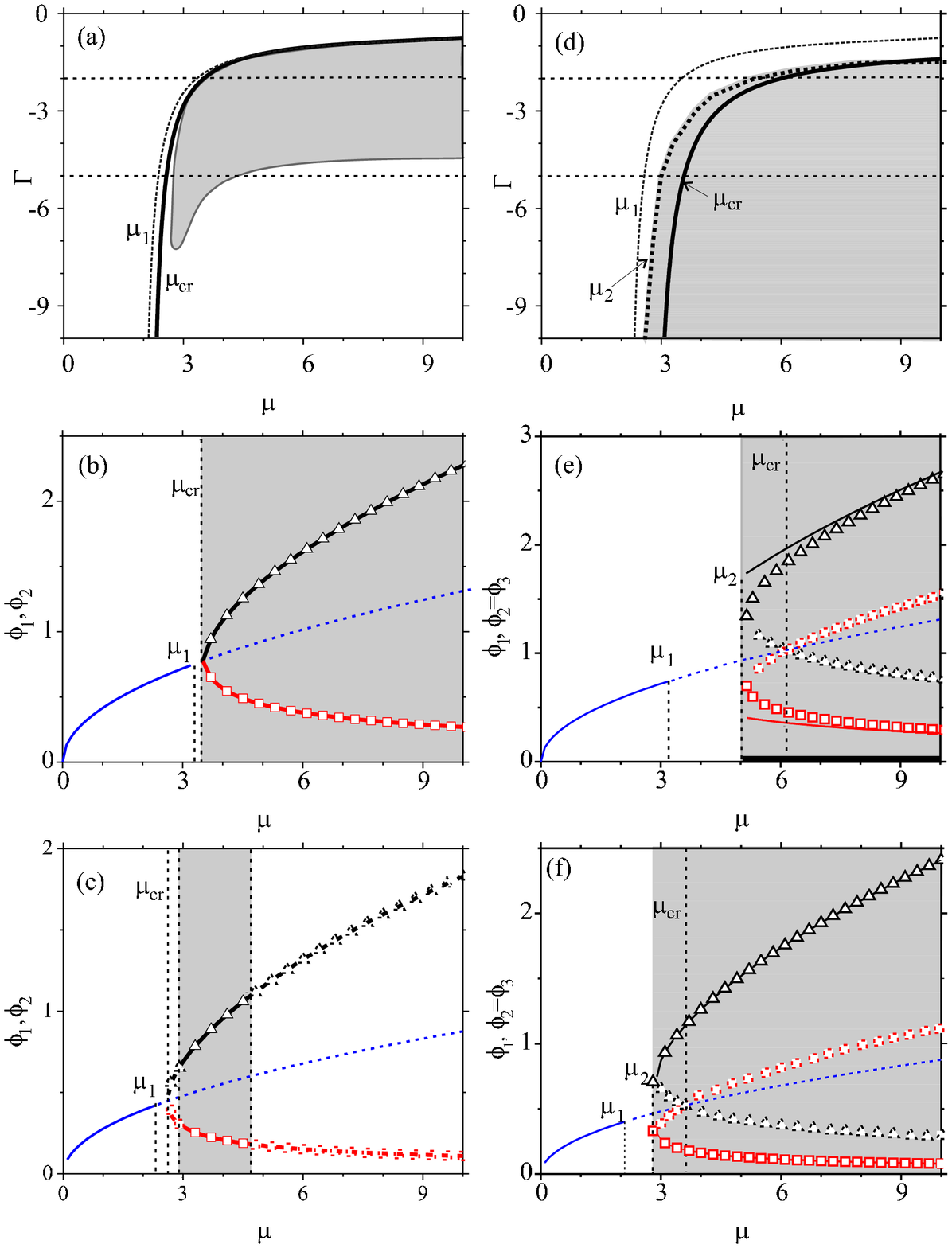} \caption{(Color
online) The stability diagram of the unstaggered DPPs (a), and
TPPs (d) for the RC+RDD interactions ($\protect\sigma =+1,\Gamma
<0$). The DPPs (a)-(c) and TPPs (d)-(f) are stable in the gray
area, while the dotted curve, $\protect\mu =\protect\mu
_{1}(\Gamma )$, separates stability and instability regions (on
its left and right sides) of CWs. The DPP is generated, via the
pitchfork bifurcation, from the CW branch by
period-doubling perturbations at $\protect\mu =\protect\mu _{\mathrm{cr}%
}(\Gamma )$, which is marked by the solid curve in (a), while the stable TPP
is generated by the subcritical pitchfork bifurcation at the turning point, $%
\protect\mu =\protect\mu _{2}<\protect\mu _{\mathrm{cr}}$ (d)-(f).
Amplitudes of the unstaggered DPP and TPP are plotted vs. $\protect\mu $ in
(b),(c), and (e), (f) for $\Gamma =-2$ and $\Gamma =-5$, respectively [these
values of $\Gamma $ are marked in (a) and (d) by dashed horizontal lines].
Chains of symbols and continuous curves show, severally, numerical results
and analytical solution for DPP (\protect\ref{eqdpp}) and TPP (\protect\ref%
{eq5}). The blue line (which seems as a bisector) represents the CW
solution. Stable and unstable states are shown by solid and dashed lines or
symbols, respectively.}
\label{dppslika}
\end{figure}

DPPs which are shown in the top row of Fig. \ref{slika1}, with $U_{n}$
taking two values, $\phi _{1}\neq \phi _{2}$, can be found as exact
solutions of Eq. (\ref{eq3}):
\begin{equation}
\phi _{1}^{2}+\phi _{2}^{2}=4\frac{\mu -2C}{4\sigma -\Gamma \zeta (3)},\quad
\phi _{1}\phi _{2}=\frac{-4C}{2\sigma +3\Gamma \zeta (3)},  \label{eqdpp}
\end{equation}%
Figure \ref{dppslika}(a) demonstrates that both the linear-stability
analysis and direct simulations reveal a rather narrow stability
area of the unstaggered DPPs for the RC+RDD sign combination of the
interactions. At large $\mu $, the upper boundary of the region is
determined by the positivity of $\phi _{1}\phi _{2}$ (i.e., the
``unstaggerness" of the pattern) in Eq. (\ref{eqdpp}) with $\sigma =+1$: $%
\Gamma <-2/(3\zeta (3))\approx -0.55$. The DPP is created by the \textit{%
supercritical} pitchfork bifurcation \cite{Iooss} (in other words, through
the phase transition of the second kind) at $\mu _{\mathrm{cr}}=2C\left[
4\Gamma \zeta (3)-2\sigma \right] /\left[ 2\sigma +3\Gamma \zeta (3)\right] $%
, which is found from Eq. (\ref{eqdpp}) by setting $\phi _{1}=\phi _{2}$. At
the bifurcation point, the DPP branch may emerge as a stable one [Fig. \ref%
{dppslika}(b)], or, at larger $|\Gamma |$, as an unstable solution, which
quickly enters the stability region with the further increase of $\mu $
[Fig. \ref{dppslika}(c)]. As seen in Fig. \ref{dppslika}(c), in the latter
case the CW solution becomes unstable against other perturbations at $\mu
=\mu _{1}$, but it remains stable against the period-doubling perturbations
up to $\mu =\mu _{\mathrm{cr}}$.

The TPPs are schematically shown in the bottom row of Fig. \ref{slika1}.
Equations for the three real amplitudes, which are designated in the figure,
are derived from Eq. (\ref{eq3}):
\begin{equation}
\mu \phi _{1}=-C(\phi _{2}+\phi _{3}-2\phi _{1})+\sigma \phi _{1}^{3}-\Gamma
\phi _{1}\left( b\phi _{1}^{2}+a(\phi _{2}^{2}+\phi _{3}^{2})\right) ,
\label{exact}
\end{equation}%
and two other equations obtained by cyclic transpositions of $\left\{
1,2,3\right\} $. Here, $a\equiv -\left[ \psi (1/3)+\psi (2/3)\right]
/54\approx 1.157$ and $b\equiv 2\zeta (3)/27\approx 0.089$, with $\psi
(z)\equiv d\left[ \ln \left( \Gamma (z)\right) \right] /dz$. These equations
give rise to three types of TPPs. \textit{Type 1} is defined by assuming $%
\phi _{2}=\phi _{3}\ll \phi _{1}$, which yields an approximate analytical
solution,
\begin{equation}
\phi _{1}=\sqrt{\frac{\mu -2C}{\sigma -\Gamma b}},\quad \phi _{2}=\phi _{3}=%
\frac{-C}{\sigma +\left( a-b\right) \Gamma }\sqrt{\frac{\sigma -\Gamma b}{%
\mu -2C}}.  \label{eq5}
\end{equation}%
\textit{Type 2} is similar to Type 1, but with comparable absolute values of
the amplitudes: $\phi _{1}>0,$ $\phi _{2}=\phi _{3}<0,\,$and \textit{Type 3 }%
is an exact solution with $\phi _{1}=0,~\phi _{2}=-\phi _{3}=\sqrt{(\mu -3C)/%
\left[ \sigma -\Gamma (a+b)\right] }$. The analysis demonstrates that solely
Type 1 may be stable, therefore only this type is considered below.

While the unstaggered TPPs exist for all combinations of the repulsive and
attractive contact and DD interactions, only the RC+RDD combination gives
rise to stable states, similar to the DPP. Again, the creation of the TPPs
is related to properties of the CW background. As seen at Fig. \ref{dppslika}%
(d), the CW becomes unstable against generic perturbations at $\mu =\mu _{1}$%
, but remains stable against the triple-period disturbances up to the
bifurcation point, $\mu =\mu _{\mathrm{cr}}$, which can be found exactly
from Eq. (\ref{exact}): $\mu _{\mathrm{cr}}=-(3C/2)\left[ \sigma
-(2a+b)\Gamma \right] /\left[ \sigma +(a-b)\Gamma \right] $. A difference
from the way the system gives rise to the stable DPP (see above) is that the
bifurcation is (weakly) \textit{subcritical }\cite{Iooss} in the present
case (alias it is the phase transition of the first kind), with the stable
TPP emerging by a jump at the turning point, $\mu =\mu _{2}<\mu _{\mathrm{cr}%
}$, see Figs. \ref{dppslika}(e),(f).

All the varieties of unstaggered DPPs and TPPs are found to be entirely
unstable for other types of the interactions, i.e., RC+ADD, AC+ADD, AC+RDD,
with the instability growth rate increasing proportionally to the lattice
coupling constant, $C$. On the other hand, at fixed $C$ the instability
growth rate \emph{decreases} with the increase of the relative strength of
the DD interactions (of either sign), which implies that they help to
increase the lifetime of unstable density waves in the dipolar BEC. Direct
simulations confirm the predictions of the linear-stability analysis, as
well as the inhibition of the instability by the DD interactions.

\begin{figure}[th]
\center\includegraphics [width=13cm]{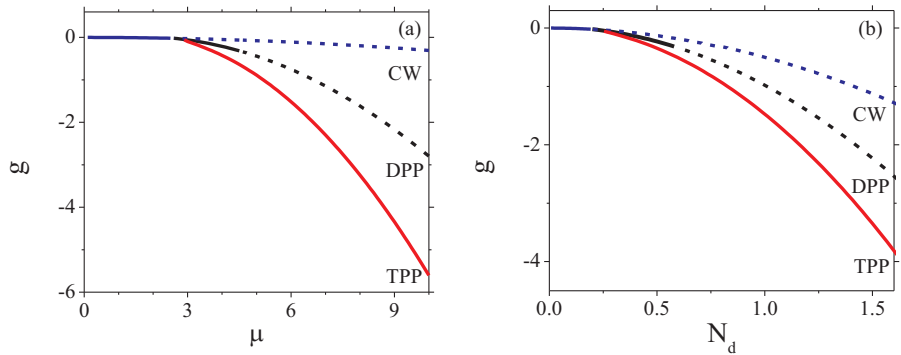}
\caption{(Color online) The free-energy density of the CW, DPP and TPP
states vs. (a) the chemical potential and (b) atomic density, $N_d$, for $%
\protect\sigma =+1$ and $\Gamma =-5$. The results are similar for other
negative values of $\Gamma $.}
\label{energy}
\end{figure}

A crucial issue is the comparison of energy of coexisting states.
Considering spatially unconfined modes, controlled by the chemical
potential, it is relevant to do this for the mean spatial density ($g$) of
the corresponding free energy, $G\equiv H-\mu N$, where $H$ and $N$ are the
Hamiltonian and number of atoms (norm of the wave function). The results,
obtained analytically for the CW and DPP, and in a numerical form for the
TPP, are shown in Fig. \ref{energy}. For a large range of parameters the TPP
is stable and realizes a relatively deep minimum of the energy, while the
DPP may only represent a metastable state. Whether the TPP is the ground
state, needs to be checked by comparing with more complex patterns, if they
exist in the system. However even if it is a metastable configuration, it is
relevant to explore its superfluid properties in further studies.

As concerns parameters relevant to the experimental implementation of the
stable condensate-density waves predicted here, in chromium the background $%
s $-wave scattering length is $a_{s}\sim 100a_{\mathrm{B}}$. By means of the
Feshbach resonance, it can be reduced to $\simeq $ $2a_{\mathrm{B}}$, thus
making the basic ratio $|\Gamma |=a_{\mathrm{DD}}/a_{s}\gtrsim 5$, which is
sufficient for getting into the stability areas displayed in Figs. \ref%
{dppslika}(a) and \ref{dppslika}(d). For the dipolar BEC of dysprosium and
erbium, the collisional properties are not yet known in detail, but similar
tuning options are expected. The creation of the DPP may not be
straightforward, as it does not correspond to the energy minimum, unlike the
TPP. To achieve this purpose, a periodic modulation of the lattice can be
applied, cf. Refs. \cite{gemelkeet, lignieret}. As concerns the observation
of the density-wave patterns, in addition to direct imaging, their existence
can be confirmed by measuring the momentum distribution in the time of
flight, after switching off the potential.

\textit{Conclusion}. Our analysis predicts stable density waves in
OL-trapped dipolar BEC, and, as a matter of fact, in other periodically
structured media with long-range interactions. In particular, the stable
TPPs are predicted for the first time, to the best of our knowledge. The
stability, which is limited to a relatively narrow region for the DPP, and
is found in a broad area for the TPP, is provided by the long-range DD
interactions, and is impossible in the BEC with solely the contact
interactions. While the DPP is metastable, the TPP realizes a relatively
deep energy minimum. These patterns emerge through the phase transitions of
the second and first kinds, respectively. These condensate-density waves are
bosonic counterparts of charge-density waves in 1D metals, created by the
Peierls instability. The results suggest that the long-range DD interactions
(whose effective strength in the quasi-1D setting can be adjusted by the
direction of the polarizing magnetic field) may be used to steer transitions
between different varieties of Mott-insulator states of the trapped bosons.
The same mechanism may generate stable density waves in other physical
settings---in particular, in optical media with nonlocal nonlinearities.

The findings reported here suggest a number of potentially interesting
developments. A straightforward question is whether condensate density waves
with periods larger than three may be stable. While this issue should be
addressed by means of an accurate analysis, a simple estimate of the ways to
minimize the energy of the system through the condensation of atoms at a
relatively small number of the lattice sites suggests that patterns with
higher values of the period may be favored by a still stronger contrast
between the strengths of the DD and contact repulsion. An extension to
multi-component condensates is interesting too. Further, Ref. \cite%
{Stuttgart}, which recently reported a stable 2D checkerboard ground state
in a related nonlocal model, suggests a possibility to search for stable
density-modulation patterns in dipolar condensates trapped in a deep 2D OL,
using the respective form of the DNLSE \cite{takode-nashi}.

On the other hand, while the analysis in this work is restricted to
the DNLSE model based on the ``frozen" lattice, the DD interactions
between condensate droplets trapped at local minima of the OL
potential may distort the droplet chain, making it necessary to add
another degree of freedom---a shift of the droplets from local
potential minima. In a combination with the equation for the
droplets' mean-field wave function, this generalization may give
rise to a new type of a dynamical system, blending the DNLSE with a
model of the Frenkel-Kontorova type. Results obtained in the
framework of such a coupled system will be reported elsewhere.
Finally, it is interesting to extend the analysis to fermionic
dipolar gases. In particular, it was very recently assumed that a
Wigner crystal may emerge in a fermionic gas trapped in the OL
\cite{China}.

\acknowledgments A.M., G.G., and Lj.H. acknowledge support from the Ministry
of Education and Science, Serbia (Project III45010). The work of B.A.M. and
T.P. was supported in a part by the German-Israel Foundation (grant No.
I-1024-2.7/2009).

\end{document}